\documentclass[aps,prl,twocolumn,groupedaddress,showpacs]{revtex4-1}
\usepackage[colorlinks=true,citecolor=blue,linkcolor=blue,pdfhighlight =/O]{hyperref}
\usepackage{graphicx}
\usepackage{amsmath}
\usepackage{color}
\usepackage{soul}
\usepackage{float}

\usepackage{natbib}

\hypersetup{ urlcolor=blue}

\begin{document}


\title{Charge Puddles in Graphene Near the Dirac Point}

\author{S. Samaddar,$^{1,2}$ I. Yudhistira,$^3$ S. Adam,$^{3,4}$ H. Courtois,$^{1,2}$ and C. B. Winkelmann$^{1,2}$}

\email{clemens.winkelmann@neel.cnrs.fr}
\affiliation{$^1$Universit\'{e} Grenoble Alpes, Institut NEEL, F-38042 Grenoble, France}
\affiliation{$^2$CNRS, Institut NEEL, F-38042 Grenoble, France}
\affiliation{$^3$Centre for Advanced 2D Materials and Department of Physics, National University of Singapore, 2 Science Drive 3, Singapore 117551}
\affiliation{$^4$Yale-NUS College, 16 College Avenue West, Singapore 138527}
\date{\today}

\begin{abstract}
The charge carrier density in graphene on a dielectric substrate such as $\mathrm{SiO_2}$ displays inhomogeneities, the so-called charge puddles. Because of the linear dispersion relation in monolayer graphene, the puddles are predicted to grow near charge neutrality, a markedly distinct property from conventional two-dimensional electron gases. By performing scanning tunneling microscopy/spectroscopy on a mesoscopic graphene device, we directly observe the puddles' growth, both in spatial extent and in amplitude, as the Dirac point is approached. Self-consistent screening theory provides a unified description of both the macroscopic transport properties and the microscopically observed charge disorder.
\end{abstract}
\maketitle

Electrons in graphene are subjected to a disordered potential created by random charged impurities, either adsorbed on the graphene or buried in the substrate. These lead to inhomogeneities in the local carrier density, that is, charge puddles \cite{Martin2007, Zhang2009, Chen2008, Tan2007, DasSarma2011}. Charge puddles are usually thought of as a limitation to the extent the charge neutrality point can be approached macroscopically, thereby also limiting possible device performances. However, the behavior of the puddles itself unveils the fascinating electronic properties of graphene and, more generally, Dirac materials. 

Electrostatic screening in two dimensions (2D) has a highly counterintuitive behavior. In conventional 2D electron gases (2DEGs), the Thomas-Fermi wave vector $q_{\rm TF}$  is independent of the carrier density $n$, meaning that the relative strength of screening, $q_{\rm TF}/k_F$, where $k_F$ is the Fermi wave-vector, increases at low densities. Because of the linear band structure in monolayer graphene and other Dirac materials, $q_{\rm TF}\propto k_F=\sqrt{\pi n}$. This has the important consequence that the unscreened potential created by a charged impurity in a medium with dielectric constant $\kappa$, $V(q)= e^2/2\kappa \epsilon_0 q$ and the screened potential ${\tilde V}(q)\propto (q+q_{\rm TF})^{-1}$ are identical within a multiplicative constant \cite{DasSarma2011}. 
In other words, near charge neutrality local inhomogeneities in the screened potential can be arbitrarily large. Further, a rough estimate of the lateral extent of charge carrier density correlations is given by $q_{\rm TF}^{-1}$, from which a strong growth $\propto n^{-1/2}$ of the puddles size is expected near charge neutrality.
The direct observation in a Dirac material of the carrier density dependence of both the charge puddles' amplitude and size has not been reported to date.

In this Letter, we report the direct microscopic observation of the doping disorder landscape in monolayer graphene at different charge carrier densities. The charge inhomogeneities are found to grow, both in spatial extent and in amplitude, as the Dirac point is approached. From transport measurements on the very same graphene sample at study, the microscopic parameters of the disorder potential can be estimated in the frame of the self-consistent screening theory. Calculations of the charge puddles distribution based on these are in very good agreement with the experimental observations.

\begin{figure}[t]
	\centering
	\includegraphics[width=\columnwidth]{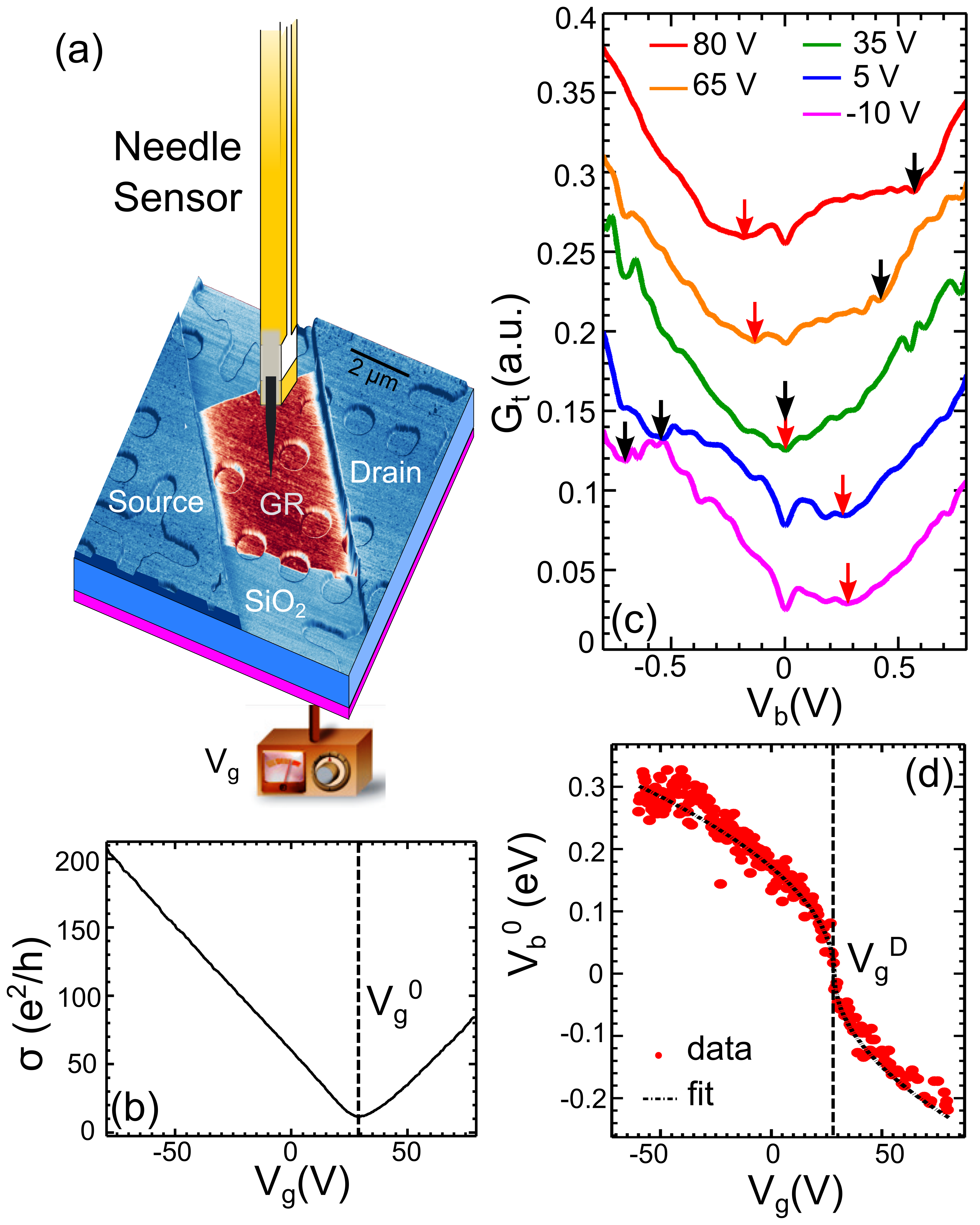}
	\linespread{1.1}
	\caption[Experimental Configuration]{(a) Experimental configuration, combining transport measurements with scanning probe microscopy on a monolayer graphene device. A conductive tip mounted on a stiff mechanical resonator serves as the probe for combined AFM-STM. Two isolated electrical contacts (\textit{Source} and \textit{Drain}) enable two-probe transport measurements. The atomic force micrograph shows both the topography (vertical scale, varying between $0$ and $57\,\mathrm{nm}$) and the phase (varying by $3.8^{\mathrm{\circ}}$ from blue to brown). (b) Device conductivity as a function of backgate voltage $V_g$, measured at a bias voltage of 5 mV. The overall charge neutrality point is found at $V_g^0 = 29\,\mathrm{V}$. (c) Differential tip-to-sample tunneling conductance $G_t$ as a function of the voltage $V_b$ (uniformly applied to the sample), at several values of $V_g$. The black and red arrows indicate the position of the primary ($V_b^0$) and the secondary minimum of $G_t$ respectively (see text). $I_t^{set}=50\,\mathrm{pA}$ at $V_b=0.9\,\mathrm{V}$. The curves are vertically offset for clarity. (d) Variation of the primary minimum with $V_g$. The black dashed line shows the fit with Eq. (\ref{Dirac}), yielding the fit parameter $V_g^D = 28\,\mathrm{V}$.}
	\label{fig1}
\end{figure}

The sample is fabricated on a heavily doped Si substrate covered with $t$=285 nm thermal oxide. A sequence of holes on the substrate surface encodes the position and allows for finding the device. Note that the holes are much shallower ($\sim 10\,\mathrm{nm}$) than their diameter ($750\,\mathrm{nm}$), thus the perturbation from this corrugation is minimal for the graphene sheet, which smoothly follows the holes' profile. Prior to exfoliation, the sample surface is cleaned with oxygen plasma. This treatment  results in a higher yield of monolayers along with a rather large density of negatively charged silanol groups  on the substrate surface \cite{Nagashio2011}. 
Single layer graphene is prepared by mechanical exfoliation \cite{Novoselov2004}. The number of graphene layers and the absence of surface contamination are confirmed from combined optical, Raman and ex-situ AFM characterization. Using a mechanical mask \cite{Choudhary2011}, we deposit the metallic source and drain contacts to form a 4 $\mu$m long graphene junction (Fig. \ref{fig1}a). Organic resist is avoided, as to ensure a residue-free surface for scanning probe microscopy. 

The experimental setup is a home-made combined AFM-STM operating at a temperature of $130\,\mathrm{mK}$ \cite{Martin2015}, at which all measurements presented here were obtained. The sample stage allows for in situ multi-terminal transport measurements of the device. AFM is performed by electrical excitation and read-out of a mechanical quartz Length Extension Resonator (LER) \cite{Heike2003, An2008}. This allows to rapidly move the tip to the graphene junction with the help of the position code \cite{Quaglio2012,leSueur2008}. Scanning tunneling micrographs reveal a clean graphene surface, following the substrate corrugation with a roughness of about $100\,\mathrm{pm}$ (see Supplemental Material file for details). Scanning tunneling spectroscopy (STS) is achieved by lock-in measurements of the differential tip-to-sample tunneling conductance $G_t(x,y)=dI_t/dV_b$. These are performed in constant tunneling current ($I_t$) mode, by adding a $12\,\mathrm{mV}$ ac modulation at frequency $f = 322\,\mathrm{Hz}$ to the bias voltage $V_b$, which is uniformly applied to the sample. 

Transport measurements are performed with the tip retracted a few hundreds of nm from the sample surface. However, approaching the tip to STM contact does not produce a significant effect in the device characteristics.
The conductivity of the graphene device shows a perfectly linear behavior at high carrier densities (Fig. \ref{fig1}b), in line with a density independent mobility of about $6000$ cm$^{2}$V$^{-1}$s$^{-1}$. This indicates that carrier transport is dominated by long range disorder, as can be caused by random charge impurities in the substrate \cite{Adam2007}. A slight difference between the measured electron and hole mobilities, $\mu_{e}/\mu_h = 0.9 \pm 0.05$  could be related to the difference in their scattering cross-sections off charged impurities \cite{Novikov2007, Chen2008}. 

The gate voltage at which conductivity is minimized gives the overall charge neutrality point, $V_g^0=29$ V. 
This overall hole doping is consistent with the presence of the negatively charged silanol groups on the surface. A residual conductivity $\sigma_0\approx 11\, e^2/h$ is found at the charge neutrality point. Within self-consistent screening theory \cite{DasSarma2011}, the above values of residual conductivity and mobility point to a charged impurity distribution with a density $n_i=7.5 \pm 0.5\times 10^{11}$ cm$^{-2}$ at a distance $0.1\, {\rm nm}<d<1\, {\rm nm}$ below the graphene, in agreement with earlier experiments in similar conditions \cite{Adam2007, Tan2007, Chen2008}.

We performed scanning tunneling spectroscopy on the graphene sheet, at distances greater than $1\,\mathrm{\mu m}$ from the metal-graphene interface as to rule out any possible influence of the leads on local properties. Fig. \ref{fig1}c shows the differential tunneling conductance $G_t(V_b)$ acquired at a given location, but at different gate voltages $V_g$. A V-shape spectrum, characteristic of graphene is obtained in every case. A frequently reported gate-independent depression of the tunneling conductance is seen at zero bias \cite{Luican2011,Jung2011,Martin2015}. In addition, the curves display two gate-dependent local minima, highlighted by red and black arrows respectively, which move in opposite directions with $V_g$. The primary minimum $V_b^0$ (red arrows) occurs when the Fermi level of the tip is aligned with the local Dirac point $E_D(\mathbf{r})$ of graphene, which can be written as:
\begin{equation}
V_b^{0} =-\gamma\,\, {\rm sign}(V_g - V_g^{D})\sqrt{|V_g - V_g^D|},
\label{Dirac}
\end{equation}
where $\gamma=\hbar v_F \sqrt{\pi \kappa \epsilon_0/(e^3 t)}$ \cite{Choudhary2011}, with $v_F=1.1\times10^{6}$ m/s the graphene Fermi velocity. The local quantity $V_g^D$ is the gate voltage at which the Fermi level of graphene is also aligned with the latter energy levels. It includes the influence of the local gating produced by the tip on the local Dirac point due to both  the tip-sample work function mismatch and the bias voltage \cite{Choudhary2011,Zhao2015}. In the absence of capacitive coupling to the tip, one would find the spatially averaged $V_g^D$ to coincide with $V_g^0$ found from transport experiments. In the case of Fig. \ref{fig1}, the experimental gate dependence of the primary minimum $V_b^0$ can be well fitted with Eq. (\ref{Dirac}), yielding $V_g^D=28$ V, see Fig. \ref{fig1}d. The value of $\gamma$ is fixed independently according to its  definition. The nearly exact matching of $V_g^D$ at that particular position and tip condition with the global value of $V_g^0$ is coincidental, since $V_g^D$ depends on the position. The secondary minimum (black arrows) occurs when the Fermi level of graphene passes through the Dirac point \cite{Jung2011,Choudhary2011,Chae2012,Zhao2015}. 
The above analysis provides a detailed understanding of the electron tunneling spectra dependence on the gate potential, at a given location. 

Several strategies can be used for mapping the local Dirac point. Performing a complete spectrum at each position, from which $E_D$ is then individually extracted, is the most reliable method but is very time-consuming \cite{Zhang2009,Martin2015}. Mapping $G_t$ at a single $V_b$, slightly offset from the average primary minimum $V_b^0$ by $\delta V_b$, was shown to reproduce qualitatively the $E_D(x,y)$ maps \cite{Zhang2009}. This stems from the fact that, to first approximation, a shift in $E_D$ simply shifts the $G_t(V_b)$ curves along the $V_b$ axis. Complications with this approach arise when one wishes to compare $E_D$ maps at different gate voltages because $V_b^0$ itself is a function of $V_g$. Our strategy consists in first determining $V_b^0(V_g)$ at a given position (Fig. \ref{fig1}d) and then mapping $G_t$ at a gate dependent bias voltage $V_b(V_g)= V_b^0(V_g) \pm 100$ mV. The sign of the offset is set such that $|V_b|>|V_b^0|$. 

We further normalize the tunnel conductance maps to the set-point tunnel conductance at $V_b$, writing ${\tilde G}_t = G_t/G_0$, with $G_0=I_t/V_b$. This normalization is known to provide a more faithful conversion of the differential tunnel conductance to the density of states, when the tunnel conductance $I_t/V_b$ is not fixed from map to map \cite{Wiesendanger1994}. We have verified the structures observed in ${\tilde G}_t$ maps to be consistent with the $E_D$ maps found from full current imaging tunneling spectroscopy (CITS) measurements, which were acquired at selected gate voltages. These full CITS also allow for determining the proportionality factor between the ${\tilde G}_t$ and $E_D$ maps (see Supplemental Material file for a more detailed discussion).

\begin{figure}[t]
	\centering
	\vspace{-0.5cm}
	\includegraphics[width=\columnwidth]{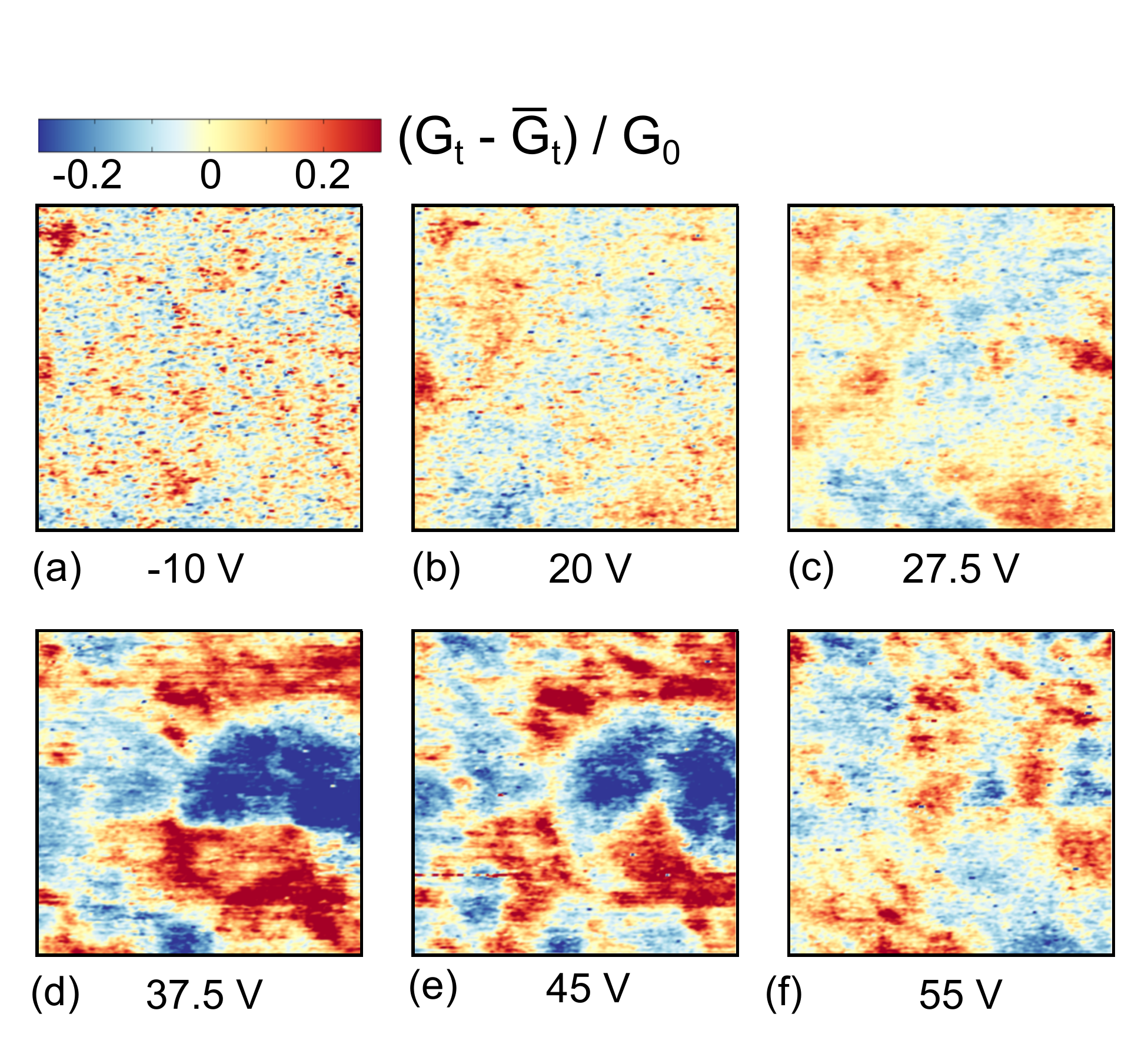}
	\linespread{1.1}
	\caption{Spatial maps of the variations of the normalized differential tunneling conductance around its mean value, over an area of $100 \times 100\,\mathrm{nm^2}$ at different gate voltages (indicated below each figure) using lock-in technique in constant current mode. The imaging parameters are a modulating voltage of amplitude $V_{\rm AC} = 12\,\mathrm{mV}$ and frequency $f = 322\,\mathrm{Hz}$ and a set-point current $I_t = 50\,\mathrm{pA}$ at a bias voltage $V_b$ equal to (a) $0.298\,\mathrm{V}$, (b) $0.191\,\mathrm{V}$ (c) $0.122\,\mathrm{V}$, (d) $-0.145\,\mathrm{V}$, (e) $-0.232\,\mathrm{V}$ and (f) $-0.267\,\mathrm{V}$.}
	\label{EHPuddles}
\end{figure}

\begin{figure}[t]
	\centering
	\includegraphics[width=0.95\columnwidth]{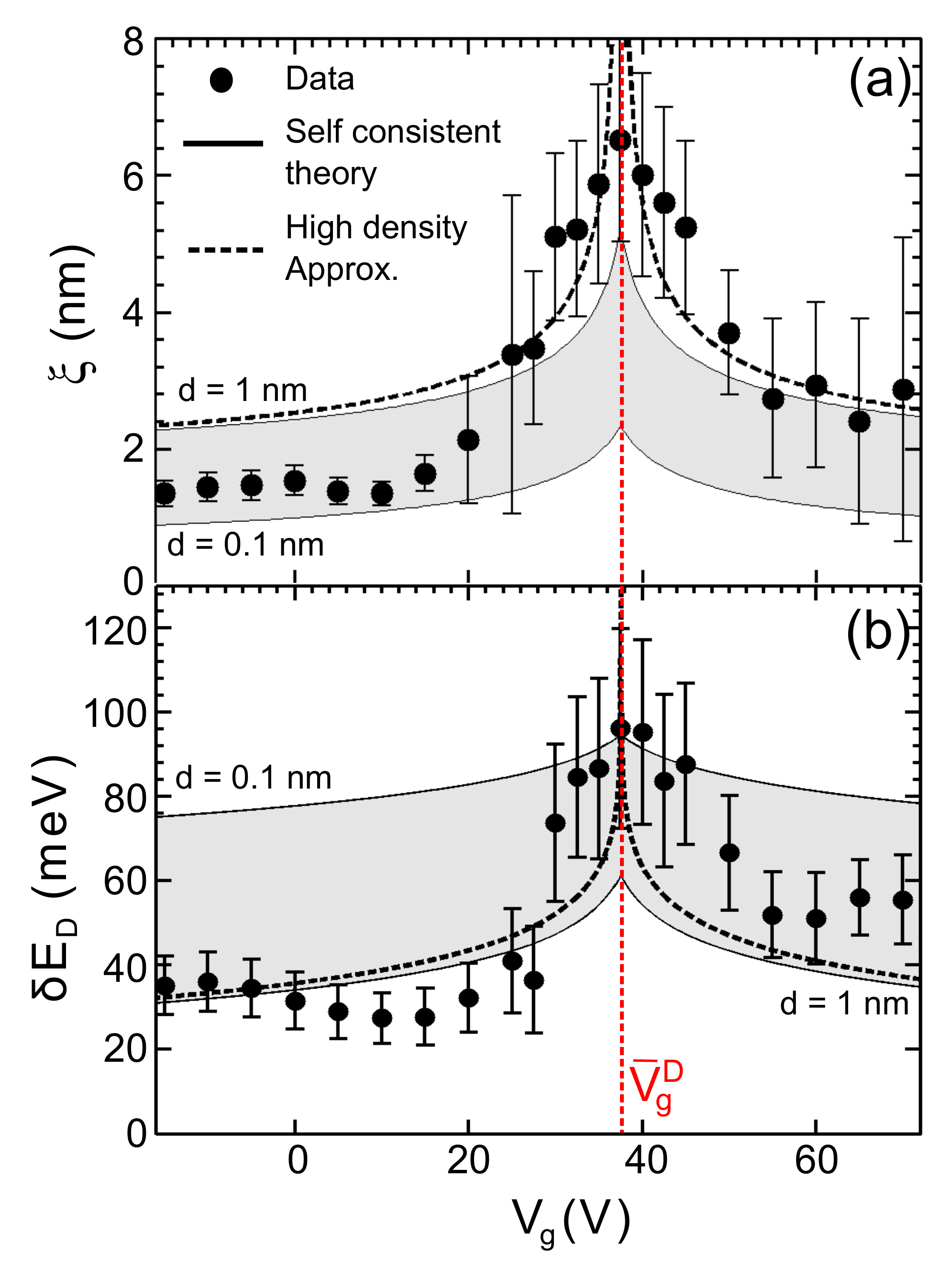}
	\linespread{1.1}
	\caption{(a) Puddle size and (b) amplitude of doping inhomogeneities, as a function of the back-gate potential. The dashed lines are calculations without self-consistent correction to carrier density, with $d = 1\,\mathrm{nm}$. The solid lines show the self-consistent screening calculations for $d = 1\,\mathrm{nm}$ and $0.1\,\mathrm{nm}$, respectively (see text). For all calculations, the value derived from transport data is used: $n_{i} = 7.5 \times 10^{11}\,\mathrm{cm^{-2}}$.}
	\label{size}
\end{figure}

Experimental  ${\tilde G}_t \sim E_D$ maps obtained with the above procedure are shown for several gate voltages in Fig. \ref{EHPuddles}. It is readily seen that not only the lateral extent, but also the amplitude of the doping inhomogeneities, gradually increase as the Dirac point is approached. For proper quantification of the observed inhomogeneities, we introduce the auto-correlation function of the $E_D$ maps. Assuming rotational symmetry (which is only approximately true, due to the finite size of the maps), the latter is  a function of only $r=|{\bf r}|$. The charge puddles' size $\xi$ is determined from fitting the angular average of the auto-correlation function of each $E_D$ map to a gaussian decay. The gate dependence of $\xi$ displayed on Fig. \ref{size}a shows a strong increase near charge neutrality, which is found at a gate voltage ${\bar V}_g^D$ of about 38 V. This value results from a spatial average over the map area. Because of the capacitive influence of the tip, ${\bar V}_g^D$ is somewhat larger than the  charge neutrality condition $V_g^0=29$ V found from transport experiments \cite{Choudhary2011,Zhao2015}. 

We further determine the amplitude of the Dirac point variations $\delta E_D$, where $\delta$ stands for the standard deviation over a map, which reflects the amplitude of the doping inhomogeneities across the sample. The amplitude $\delta E_D$ is plotted on Fig. \ref{size}b as a function of $V_g$ and also shows a marked peak at ${\bar V}_g^D \approx 38$ V. The error bars on $\xi$ and $\delta E_D$ are mainly associated to the finite size of the maps; a detailed discussion of their determination can be found in the Supplemental Information file. Some asymmetry of the puddles' behavior is observed, which appear somewhat larger and more prominent at large electron doping, than on the hole doped side. As electron doping involves a quite large gate potential of about $60$ V, a possible scenario for this asymmetry is that the back-gate eventually influences the substrate impurities distribution itself \cite{Tan2007}. 

Our main experimental findings are thus that both the amplitude and the spatial extent of the puddles significantly increase as the Dirac point is approached. For a quantitative understanding, we now compare these results to calculations. From Thomas-Fermi theory in 2D, assuming a flat Fermi see, follows that local variations in the local value of $E_D/e$ are equal to variations in the screened electrostatic potential $\tilde V$ \cite{Ashcroft}. For a random distribution of charged impurities with density $n_i$ at a distance $d$ from the graphene sheet, the auto-correlation function of the screened potential
can be written \cite{Adam2011}
\begin{equation}
C(r) = 2\pi n_i \left(\frac{e^2}{4 \pi \epsilon_0 \kappa}\right)^2 \int_0^{+ \infty} q\, dq \, \left[ \frac{1}{\epsilon(q)}\frac{e^{-qd}}{q} \right]^2 J_0(qr),
\label{C}
\end{equation}
where $J_0$ is the zeroth-order Bessel function and $\epsilon(q)$ is the graphene dielectric function. The latter describes the screening of Dirac fermions which, within Random Phase Approximation (RPA), can be written as \cite{Adam2007}
\begin{equation}
\epsilon(q)=
\begin{cases}
	1+ 4 k_F r_s/q  & \text{ for $q< 2k_F$,} \\
	1+ \pi r_s/2  & \text{ for $q > 2k_F$,}
\end{cases}
\label{rpa}
\end{equation}
where $r_s=e^2/(4\pi \epsilon_0 \kappa \hbar v_F)\approx0.8$ on SiO$_2$ is the effective fine structure constant of graphene. The dependence of the correlation function on the mean doping level (and thus on the gate potential) enters here through the dependence of $\epsilon(q)$ on $2k_F$.

We calculated the auto-correlation function for the screened potential and extracted the correlation length $\xi$ and fluctuation amplitude $\delta \tilde V = \delta E_D/e$. The result for $d=1$ nm, shown as dashed curves in Fig. \ref{size}a,b, accounts for the overall decrease of both $\xi$ and $\delta E_D$, that is, stronger screening, with increasing charge carrier density. The calculations have no other adjustable parameter than ${\bar V}_g^D = 38 V$, the impurity density in the substrate $n_i$ being determined from the transport measurements. The puddles size follows in particular the expected trend $\xi \sim q_{TF}^{-1} \propto n^{-1/2}$ at high carrier densities, where $n\propto |V_g-{\bar V}_g^D|$ is the gate induced charge carrier density. This agreement validates the microscopic picture of random potential fluctuations, for the description of which we call for Thomas-Fermi screening in a Dirac material.

At charge neutrality and for a homogeneous system, there are no excess charges available to screen the impurity potential. Accordingly, Eq. (\ref{rpa}) predicts that both the amplitude and the correlation length diverge. However, this ignores the fact that the induced charges within the puddles can themselves screen the impurity potential. Accounting for this process self-consistently \cite{Adam2007} leads to rewriting the RPA polarizability of Eq. (\ref{rpa}) with a corrected charge carrier density. The usual expression of $k_F=\sqrt{\pi n}$ is then replaced by $\sqrt{\pi (n+n^*)}$ \cite{Adam2011}, where $n^*$ represents the disorder-induced charge carrier density which cannot be compensated by a global gate.
The self-consistent calculations are plotted in Fig. \ref{size}a,b. The grey regions are comprised between the theoretical curves for $d = 0.1$ nm and $1$ nm respectively. The ensuing saturation of both the puddles' size and amplitude at the charge neutrality point is in very good agreement with the experimentally observed trend. 

To conclude, this work provides the first microscopic observation of the growth of charge inhomogeneities in graphene near the Dirac point. It further shows that the observed behavior can be very well described with a theory based on a microscopic description of the impurity potential, using parameters found from transport measurements, performed {\it in situ} on the very same graphene sample. This observation gives utmost credit to the charged impurity potential scenario as a limiting factor to the exploitability of Dirac physics in graphene. 

This work was funded by the European Commission under project no. 264034 (Q-NET Marie Curie Initial Training Network). Work in Singapore was supported by the National Research Foundation of Singapore under its Fellowship programme (NRF-NRFF2012-01). Samples were fabricated at the Nanofab facility at Institut N\'eel. We thank S. C. Martin, B. Sac\'ep\'e, A. de Cecco, J. Seidemann and A. K. Gupta for discussions and help with the experiments. 

\bibliography{./Bibliography}
\end{document}